\documentclass[%
 aip,
 amsmath,amssymb,
 reprint,%
]{revtex4-1}

\usepackage{graphicx}
\usepackage{dcolumn}
\usepackage{bm}

\usepackage[utf8]{inputenc}
\usepackage[T1]{fontenc}
\usepackage{mathptmx}
\usepackage{xcolor}
\usepackage{soul}
\usepackage{pgffor} 
\usepackage{pdfpages} 
\makeatletter
\AtBeginDocument{\let\LS@rot\@undefined}
\makeatother

\begin{document}

\preprint{AIP/123-QED}

\title[]{Sound absorption in Hilbert Fractal and Coiled Acoustic Metamaterials}

\author{G. Comandini}
\email{gianni.comandini@bristol.ac.uk}
\affiliation{School of Civil, Aerospace and Mechanical Engineering (CAME), Bristol Composite Institute (BCI) University of Bristol.}

\author{C. Khodr}
 \altaffiliation[Currently addressed at ]{Ecole Centrale de Lyon ECL Laboratoire de Mécanique des Fluides et Acoustique (LMFA)}

\affiliation{School of Civil, Aerospace and Mechanical Engineering (CAME), Department of Mechanical Engineering, University of Bristol.}

\author{V. P. Ting}
\affiliation{School of Civil, Aerospace and Mechanical Engineering (CAME), Bristol Composite Institute (BCI) University of Bristol.}

\author{M. Azarpeyvand}
\affiliation{School of Civil, Aerospace and Mechanical Engineering (CAME), Department of Aerospace Engineering, University of Bristol.}

\author{F. Scarpa}
\affiliation{School of Civil, Aerospace and Mechanical Engineering (CAME), Bristol Composite Institute (BCI) University of Bristol.}

\date{\today}

\begin{abstract}
We describe here a class of acoustic metamaterials with fractal Hilbert space-filling and coiled geometry with equal tortuosity for noise mitigation. Experiments are performed using a four-microphone impedance tube and benchmarked against non-viscous and viscothermal Finite Element models related to configurations spanning up to five fractal/geometry orders. We show that the acoustic absorption can be predicted by the resonance of the cavities associated to the tortuous paths. For a given fractal/geometry order, the acoustic absorption at specific frequencies is also enhanced by maximising the difference between the minimum and maximum fluid particle velocity of the air inside the patterns. These principles can be used to design high-performance acoustic metamaterials for sound absorption over broad frequency ranges. 
\end{abstract}

\maketitle

Rapid developments in the field of acoustic metamaterials has led to the design of new solutions for mitigating broadband noise \cite{fang2017ultra}. Unlike classical absorbers, acoustic metamaterials rely on repeated/periodic sub-wavelength structures that modify the phase $v_{\phi}$ and group $v_g$ velocities of sound\cite{jimenez2016ultra,li2016simultaneous}. It is now possible to design metamaterials with null or negative density $\rho$ and bulk modulus $\kappa$\cite{gracia2013negative, cummer2016controlling}, and those anomalous properties are associated with wave phenomena such as acoustic cloaking\cite{fan2020reconfigurable,basiri2021non} ($\rho=0$), super-lenses\cite{park2015acoustic,iyer2008mechanisms} ($\rho$, $\kappa<0$) and sound slowness\cite{li2013unidirectional,ma2015dispersion} ($\kappa=0$). Labyrinth metamaterials have previously shown promising results for noise management applications\cite{xie2013tapered,kumar2020labyrinthine}. Here, we describe the acoustic properties of 3D fractal-shaped metamaterials (MMs)\cite{song2016broadband,grigorenko2013nanostructures}, previously used only for applications involving the space-coiling of electromagnetic waves\cite{xie2017microwave,wang2011experiments}. The description is made via impedance tube experiments and numerical simulations. The space-filling curve used in this work is the Hilbert fractal\cite{zhao2018fractal,man20193d} (Fig.~\ref{fig:uno}c, and Fig.~\ref{fig:uno}e). The results are benchmarked against those from a non-fractal coiled geometry\cite{lawder2000application} (Fig.~\ref{fig:uno}d, and Fig.~\ref{fig:uno}f). The patterns of the two configurations possess same lengths and internal volumes, with a minimum number of $90$ degrees angled paths for the coiled geometry (see Fig.~\ref{fig:uno}h and Fig.~\ref{fig:uno}i, generated from the same area subdivision in Fig.~\ref{fig:uno}g,  and SI).
The fractal MMs are 3D-printed with  polylactic acid (PLA) using an Ultimaker 2+ machine\cite{hernandez2015factors}. The percentage of the PLA infill during the  printing was kept constant at 30\%. 

\begin{figure}[b!]
\includegraphics{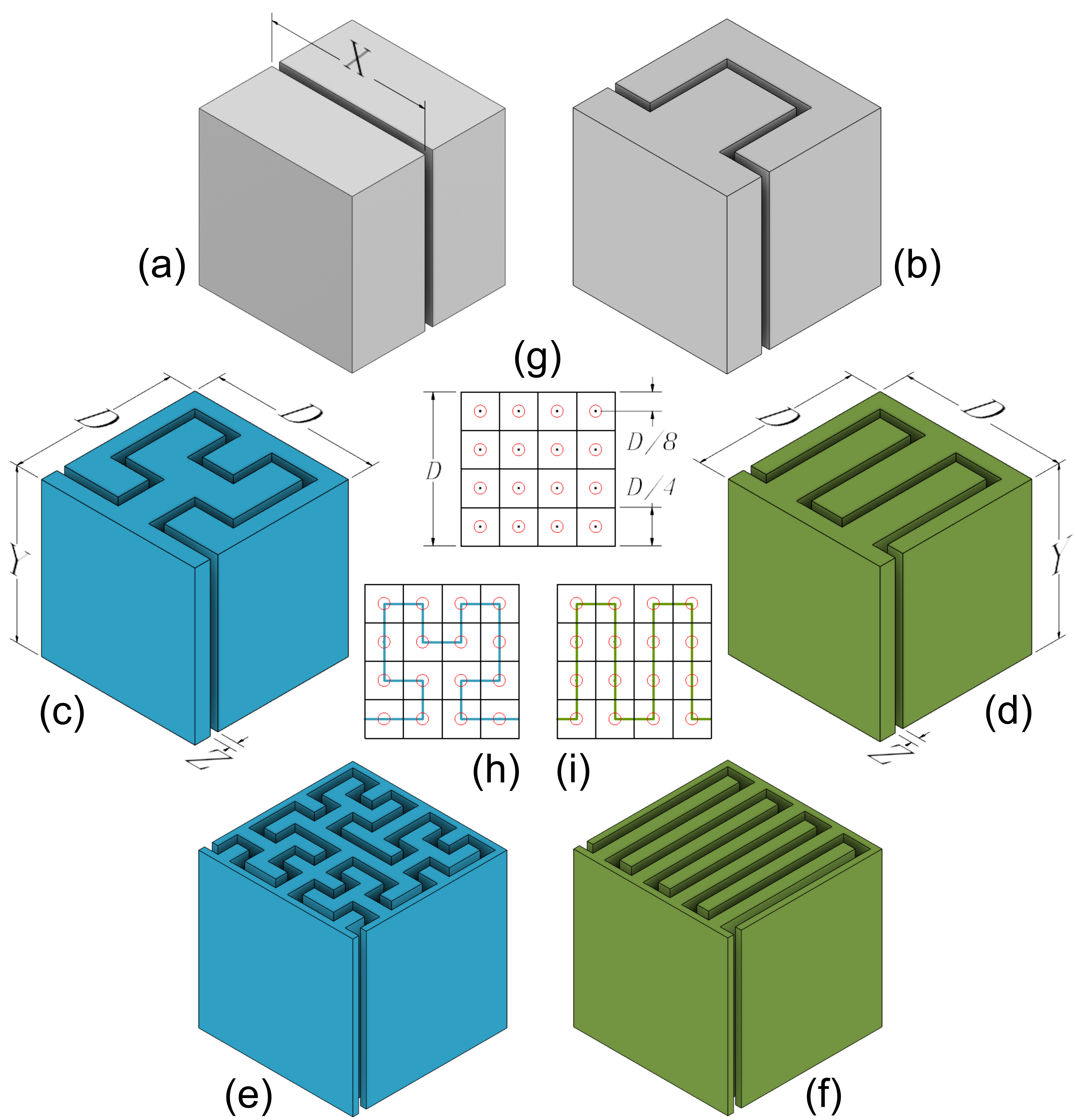}
\caption{\label{fig:uno} 
(c) Second order, Hilbert fractal. (d) Second order, coiled geometry. (a, b) Orders zero and first, common for the two configurations. (e) Third order Hilbert. (f) Third order coiled geometry. (g) The design process of the two metamaterials involves a subdivision of the square area in equal portions with relative centres. Two distinctive connective paths of the central dots will create a different space filling curve, like the (h) Hilbert or (i) the coiled one.}
\end{figure}

\begin{figure}
\includegraphics{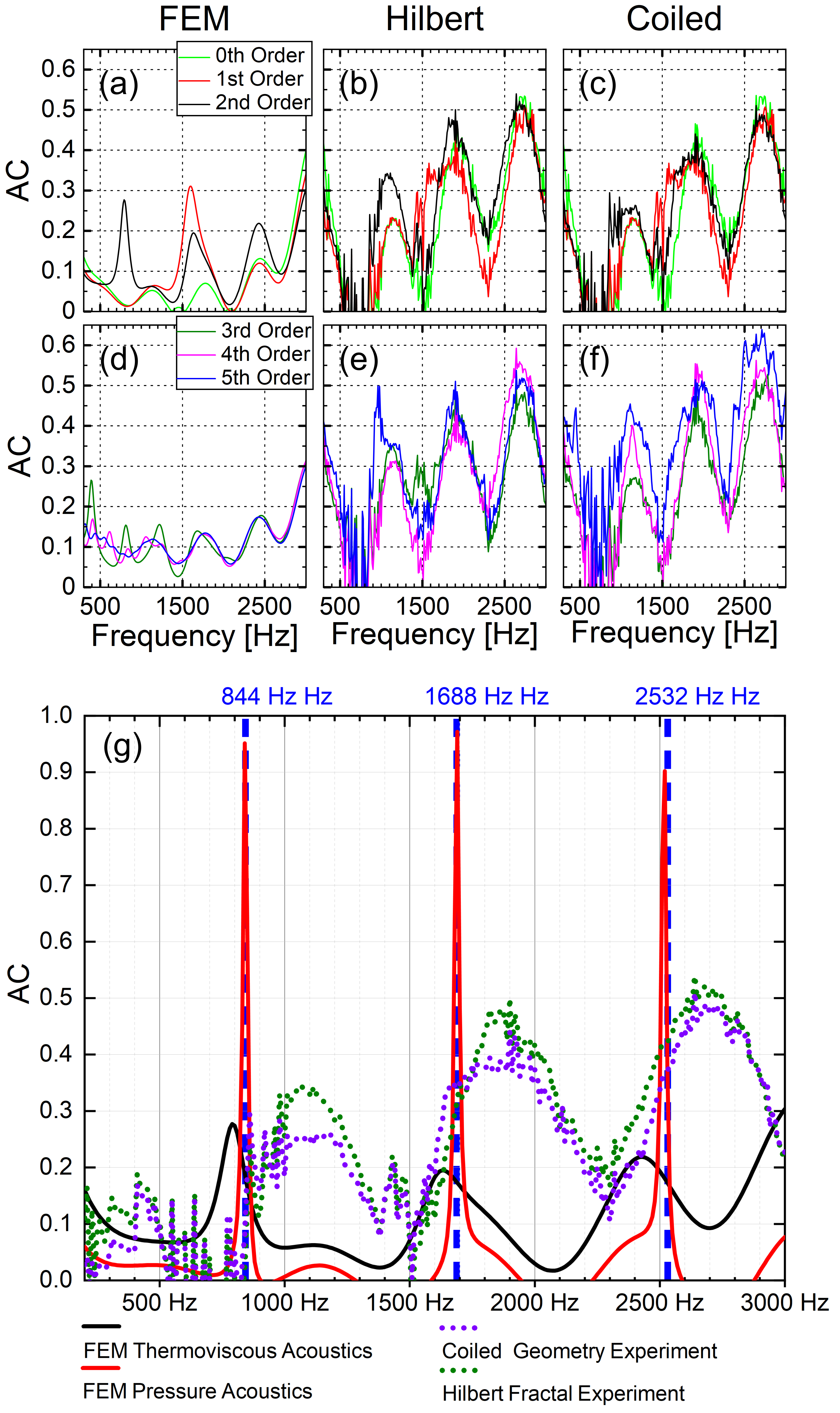}
\caption{\label{fig:due} (a) FE Absorption coefficients for the zeroth, first and second-order Hilbert fractal and coiled geometry; the latter  coincide in the numerical models.  Experimental results for the Hilbert (b) and coiled (c) from the zeroth to the second order.
Absorption coefficient for the third, fourth and fifth-order of the two geometries. (d) FE model of both configurations. Experimental (e) Hilbert and coiled (f). (g) Absorption coefficients for the second Hilbert and coiled patterns related to experiments, FE viscous and non-viscous models. Vertical lines are also drawn to indicate the first 3 resonance modes of the patterned cavities. }
\end{figure}

The Hilbert MM fractal patterns here range from the zeroth (Fig.~\ref{fig:uno}a) to the fifth order, with gap widths (indicated as Z in Fig.~\ref{fig:uno}c, and Fig.~\ref{fig:uno}d) between 0.5 mm and 3.0 mm. The maximum fractal order and gaps dimensions are constrained by the manufacturing capability of the printer. The acoustic absorption (AC) has been measured following the ASTM E2611--09 standard\cite{astm2009astm} by using a four-microphones impedance tube (more details in the Supplementary Information). The frequency bandwidth considered here is within the 0.2 kHz - 3 kHz range. Each measurement has been repeated ten times to remove outliers, using the Chauvenet's criterion with Normal distribution and 50\% threshold\cite{lin2007cleaning}. 
 For each fractal and coiled geometry we have also computed the resonance associated to an equivalent rectangular section that represents the labyrinth path with both ends open. We have used here  Eq.~(\ref{eq:1}), where \textit{c} is the air speed of sound, \textit{i, j, k} are the mode numbers, \textit{X} the fractal length (Fig.~\ref{fig:uno}a) as calculated in Eq.~(\ref{eq:1}), \textit{D} is the external dimension (Figs.~\ref{fig:uno}c, d, g) of the single MM (50.8 mm in this case), \textit{Y} and \textit{Z} are the height and gap width of the MM (Fig.~\ref{fig:uno}c and Fig.~\ref{fig:uno}d).
 
\begin{equation}
f=\frac{c}{2}\sqrt{\left ( \frac{i^{2}}{X^{2}}+\frac{j^{2}}{Y^{2}}+\frac{k^{2}}{Z^{2}}\right )}
\label{eq:1}
\end{equation}

 Finite Element (FE) simulations of the cavity modes with and without air viscosity have also been carried out using COMSOL Multiphysics. The analytical resonance frequencies are in good agreement with the ones simulated via FE (see Tables S1 and S2), with average differences ranging from 0.15\% and 0.9\% between the 5th and the 1st order. The acoustic absorption has also been simulated with full-scale FE models  (more info on the FE in the SI). 

\maketitle

Fig. \ref{fig:due} shows the direct comparison between the  measured and simulated acoustic absorption of the fractal and coiled patterns at different orders. The absorption results show a resonance-type behaviour, with peaks depending upon the order of the fractal or coiled geometry. Good agreement can be observed between numerical and experimental AC values in terms of trends, with the experimental ones tending to increase toward values of 0.5 at frequencies beyond 1.6 kHz. This behaviour is remarkable, when one considers that these metamaterials have equivalent porosity between 2\% and 63\% only (see SI). Some differences between the FEM (Fig.~\ref{fig:due}a, and Fig.~\ref{fig:due}d) and experimental results  (Fig.~\ref{fig:due}b and Fig.~\ref{fig:due}e for Hilbert and Fig.~\ref{fig:due}c and Fig.~\ref{fig:due}f for the coiled) are observed for the acoustic absorption related to zeroth, first and second order, as well as for the third, fourth and fifth. 
Differences can be ascribed to manufacturing imperfections related to the fusion deposition moulding and local slack in the test rig. The viscous FE results however mirror the periodic arrangement of the experimental absorption peaks; the experimental frequencies associated to the main peaks are larger than the numerical ones ({\char`\~}30\% on average). The viscous FE tends to underestimate the amplitude of the AC peaks, although at lower frequencies the comparison is good (0.28 against 0.33 for the first peak of the 2nd Hilbert fractal - see \ref{fig:due}g). The AC peaks correspond to fundamental cavity modes of the fractal pattern along the $X$ direction, where X=$D(2^n)$ due to Helmholtz resonators effects ( Fig.~\ref{fig:uno}c and Fig.~\ref{fig:uno}d). One can also observe the presence of small discrepancies between the resonances of the pattern cavities predicted by Eq.~(\ref{eq:1}) and the non-viscous FE model (between 0.09\% and 0.15\% - see Tables SI and SII). The frequencies corresponding to the peaks in the AC in Fig.~\ref{fig:due}a and Fig.~\ref{fig:due}d are related to resonant phenomena inside the air ducts. Quite importantly, no difference of acoustic absorption is observed between the FE-simulated Hilbert and coiled patterns, no matter which fractal order is considered. This behaviour is also substantially confirmed by the experiments, both in terms of AC peaks frequencies and amplitudes (Fig.~\ref{fig:due}b, \ref{fig:due}c, \ref{fig:due}e and \ref{fig:due}f). The Hilbert fractal and the coiled geometry have a different number of 90$^{\circ}$ angles in their geometry (Fig. S1 in supplementary information) and consequently different internal  numbers of right angles forming the Hilbert fractal. It is however apparent that the AC behaviour globally depends on the overall tortuosity, and not by the number of ninety-degree angles forming the simple fractal polygonal chain.   

\maketitle

\begin{figure}
\includegraphics{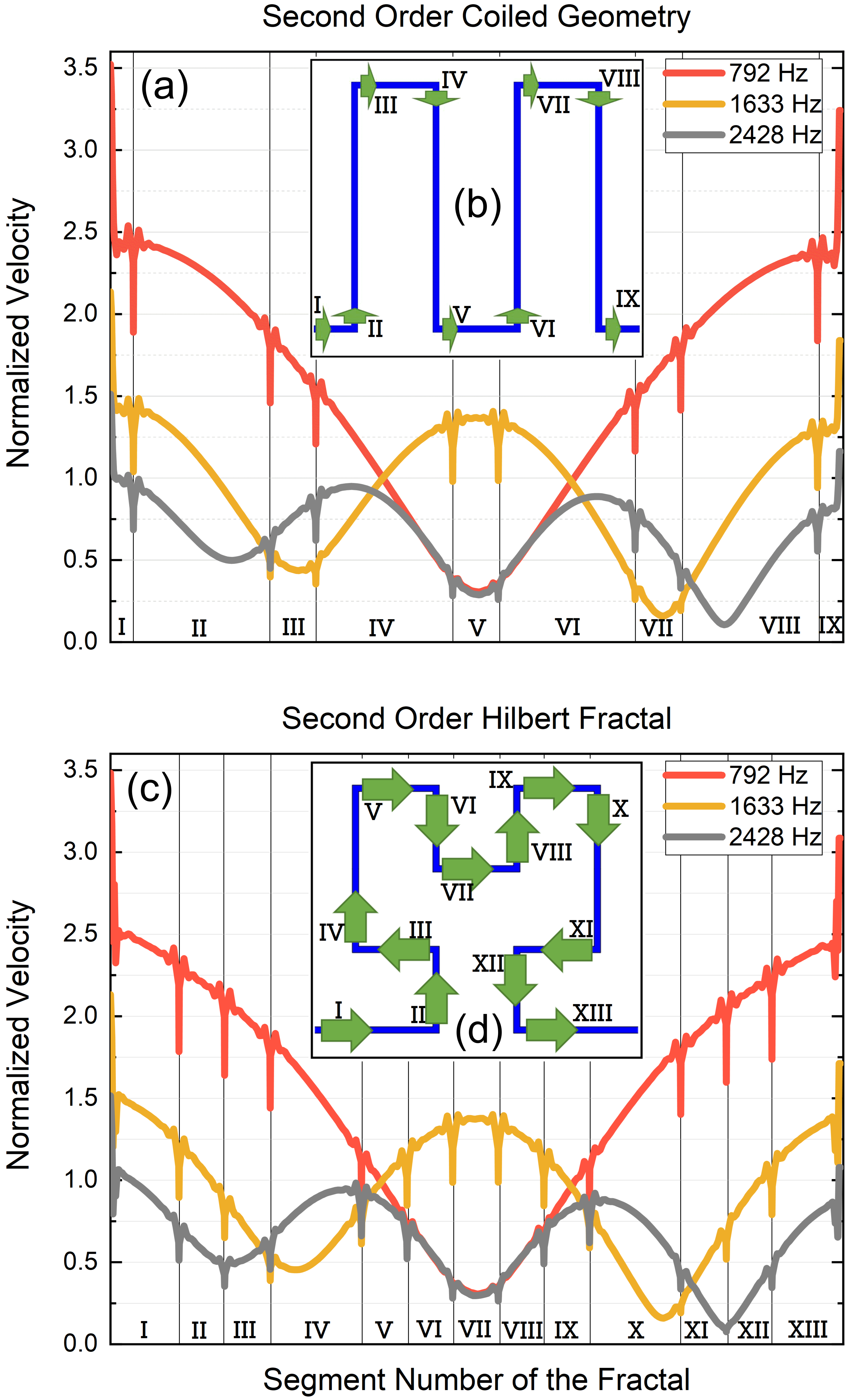}
\caption{\label{fig:tre} (a) Calculated distribution of the RMS viscous FE velocity inside the second-order coiled geometry, normalised with the RMS value of the velocity at the inlet of the first order Hilbert fractal (5.6 mm/s). The velocity graph refers to AC peaks in (Fig.~\ref{fig:tre}b) at 792 Hz, 1633 Hz and 2428 Hz. Drops in velocity coincide with losses due at the $90^o$ angles of the patterns. (b) Locations and directions of the velocity field. Every roman number identifies a segment of the geometry where the velocity has been calculated. (c) RMS local velocity field corresponding to the three AC peaks for the second Hilbert fractal between 200 Hz and 3000 Hz of (Fig.~\ref{fig:tre}d). (d) Position and direction of the calculated velocity field in the Hilbert fractal, second-order.}
\end{figure}

Not all the acoustic cavity modes maximise the absorption (see for example the FE in Figs.~\ref{fig:due}a, \ref{fig:due}d, and Figs. SI2 and SI3). To clarify this aspect, we have calculated the distribution of the root mean square (RMS) of the normalised local velocity inside the tortuous path of the two 2nd order metamaterials between 0.2 kHz and 3 kHz (Fig.~\ref{fig:tre}b, and Fig.~\ref{fig:tre}d). The maximum of the absorption peaks correspond to the largest differences in normalised velocity between minimum and maximum RMS values. 
This can be observed by looking at the AC peak at 1633 Hz of second order Hilbert fractal (Fig.~\ref{fig:tre}d), which coincides with the maximum difference between min and max normalised velocity at the middle of the tortuous path
(Fig.~\ref{fig:tre}d). At the other frequencies corresponding to the AC FE peaks (792 Hz and 2428 Hz) the velocity field shows the presence of two and three minima with lower maximum difference (0.5 and 0.8). A very similar behaviour is observed for the coiled pattern of the same order (Fig.~\ref{fig:tre}a). The number of right angles forming the two polygonal geometries is different, but the normalised velocity field is similar for the two geometries (Fig.~\ref{fig:tre}a and Fig.~\ref{fig:tre}c).  Another noticeable point is that the behaviour of the AC peaks and maximum difference in internal velocities both follow the fractal order of the geometries, in terms of number of peaks and valleys, keeping the maximum difference between the lower and higher values of the RMS velocity, only in correspondence of the main AC peaks.
 
\maketitle

Fractal acoustic metamaterials can be effectively used to tailor the acoustic absorption over a broad frequency range by designing the pattern with specific tortuous path lengths and depth of the slit. The absorption is provided through a series of multiple peaks that depend on the fractal or geometric order of the patterns. The frequencies corresponding to those peaks can be predicted quite well by the resonances of non-viscous fluid cavities with overall equal path length and depth of the patterns. Finite Elements that include viscothermal effects simulate peaks corresponding to slightly lower frequencies and trends in broad agreement with experimental data, although the latter show higher AC values. For a given fractal/geometry order, the maximum values of the absorption correspond to particle air velocity fields inside the tortuous paths that feature the largest difference between minimum and maximum velocity values. No substantial difference in AC behaviour is observed between the fractal Hilbert and the coiled geometry: although with different $90^o$ values, the two configurations have the same tortuosity \textit{X/D}. The fractal Hilbert geometry is, however, a self-filling space by definition: this pattern can therefore be a preferred choice to design space-optimised fractal geometries to enhance the acoustic behaviour via metamaterials.    

\maketitle

GC acknowledges the support of UK EPSRC through the ACCIS Composites Centre for Doctoral Training, and Ali Kandemir, Jibran Yousafzai, Alper Celik and Abhishek Gautam for technical assistance. VT acknowledges funding from (EP/R01650X/1). FS acknowledges the support of ERC-2020-AdG-NEUROMETA (101020715).

\nocite{*}
\bibliography{aipsamp}


\end{document}